\documentclass{article}
\usepackage[utf8]{inputenc}
\usepackage{graphicx}
\usepackage[utf8]{inputenc} 
\usepackage[T1]{fontenc}    
\usepackage{hyperref}       
\usepackage{url}            
\usepackage{booktabs}       
\usepackage{amsfonts}       
\usepackage{nicefrac}       
\usepackage{microtype}      
\usepackage{amsmath}
\usepackage{xcolor}
\usepackage{graphicx}
\usepackage{adjustbox}

\title{Combating Anti-Blackness in the AI Community}
\author{
  Devin Guillory \\
  University of California - Berkeley\\
  \texttt{dguillory@cs.berkeley.edu} \\
  }

\begin{document}

\maketitle

\begin{abstract}
In response to a national and international awakening on the issues of anti-Blackness and systemic discrimination, we have penned this piece to serve as a resource for allies in the AI community who are wondering how they can more effectively engage with dismantling racist systems. This work aims to help elucidate areas where the AI community actively and passively contributes to anti-Blackness and offers actionable items on ways to reduce harm.  
\end{abstract}

\section{Introduction}
 ``How did you go bankrupt?!'' Bill asked. ``Two ways,'' Mike said. ``Gradually and then suddenly'' -- Hemmingway \cite{hemingway1926sun}. This oft referenced phrase aptly describes how substantial changes that take a long time to develop can appear to happen all at once. The extrajudicial killings of George Floyd, Breanna Taylor, Ahmed Arbrey, Tony McDade, and others, in combination with a poorly-managed pandemic that is disproportionately damaging the Black community, has poured gas on a growing blaze, calling on us to truly address the extent of anti-Black systemic racism in the United States and abroad. Through this lens, many in the AI community are asking: ``What can I do to combat systemic racial injustice?". The aim of this work is to help community members better identify and understand the scale and scope of anti-Black bias within our AI community and illustrate some concrete steps that members can take to help mitigate these issues and build a more just community.
 
 To summarize our contributions we first establish the necessity of recognizing the scale and scope of anti-Blackness and how it permeates all of our institutions. We then identify areas within academia where anti-Blackness is magnified or reinforced and propose actions for faculty, graduate students, and conferences to take to minimize these deficiencies.

 \section{Background}
 
Students of optimization go through a stage where they see every problem as an optimization problem. Similarly, students of racism must go through a stage where they realize that anti-Black racism impacts every aspect of our society. This acknowledgement is not sufficient to dismantle systems of racism; however, it is a necessary first step and allows one to better appreciate the scope of the issues that we are trying to solve. Issues such as access to healthcare\cite{feagin2014systemic}, clean air \cite{bravo2016racial}, quality education \cite{harper2009access}, credit \cite{blanchflower2003discrimination}, clean water \cite{pulido2016flint}, housing \cite{williams2005changing}, public transit \cite{williams2005changing}, voter suppression \cite{bentele2013jim}, fair wages \cite{coleman2003job}, and criminality \cite{brewer2008racialization} all have roots in anti-Blackness.


 
\par
We restrict our focus to the AI community and ways in which we contribute to anti-Black racism. While the focus of recent protests have been centered around anti-Black bias in the criminal justice system, we will expand our scope to include more ways in which we propagate systemic harm: in our sources of funding, in who we allow to participate in our community, in what problems we address or exacerbate through our research and applications, and in what set of principles we allow to guide us.

\section{Examining Systems of Racism}
We propose a simplifying model for examining the sources of systemic racism that permeate our communities. The first source of systemic racism that we explore are discrepancies in physical resources. This includes differences in wealth, income, access to computing resources, access to clean air or water, healthcare, and transportation, amongst other areas. Included in this list, though perhaps more abstract, is time; time spent dealing with all other issues of race is also time not explicitly spent improving skills that are more greatly valued than "the ability to navigate racism." The second source that we will explore is social resources. Who we know and who knows us are tremendous factors in explaining the opportunities presented in our life \cite{calvo2004effects}. Large percentages of open roles in technology companies are filled via referrals; the compensation structures in CS/AI are hidden and structured in such a way that you have to know someone to know what is fair. Projects are rarely, if ever, completed with one individual and who you know informs what you work on, what opportunities you are aware of, who can vouch for you, and what social structures accept you as a member. Jones et al.(2013) \cite{jones20132013, jones2014self} observe that $75\%$ of White people have entirely White social networks, while the average White persons social network is only $1\%$ Black. This remarkable racial stratification of social networks combined with the prominent role that social networks play in our successes, lead to further racial inequity. Lastly, we look at racial discrepancies in ``measures'', i.e., anything that is used to evaluate, punish, or reward individuals. At this moment, the most prominent example of measures are discrepancies in policing. While Black and White people use drugs at similar rates, those arrested and convicted for drug offenses are disproportionately Black. Measures of aptitude, such as the SAT or GRE, have also shown racial bias. The same can be said for teacher evaluations, coding interviews, open-source project contributions, company promotion cycles, and in-school suspension rates. A overwhelming number of measures of quality have been shown to exhibit or allow for racial biases. These three categories of systemic bias--physical resources, social resources, and measures--are not all encompassing and many overlap (e.g., credit may be viewed as a measure of quality or a physical resource), but they serve as a good starting point to dissect systems of systemic bias.


\section{Academia}

When examining systems, whether they be systems of injustice and otherwise, the importance of feedback mechanisms cannot be understated. Since Academia serves as both the touchstone of the AI community as well as the developmental environment for new AI talent, any biases or discrepancies within Academia will be propagated to other places that AI touches and feed back into the next generations of Academia where it may further enhance systemic biases. As such, Academia is not only obligated to halt practices that disproportionately impact the Black community, but also to repair the damage that has been done through generations of neglect.

\subsection{Faculty}

Our faculty set the tone for our community. If our community is to ever achieve equity, it will inevitably require buy-in and focused effort from our faculty. Our professors decide which topics and prior research are highlighted in courses, who is admitted into our programs, and what criteria is required to attain a degree. They are tasked with simultaneously advancing the state of our field and training the next generation of leaders. Consequently, it is pertinent for AI faculty to fully understand all the ways that systemic racism pervades our field and help identify ways to address it.  

\begin{itemize}
    \item Who is admitted?
    \par
    We assert that there is no objective way to determine research potential, and we believe that this is the primary goal of the admissions process. Grade point averages, letters of recommendation, research publications, and GRE scores are some of the major factors used in determining who will be admitted to a graduate program. Studies have shown that GRE scores are racially biased \cite{miller2014test}. Publications and letters of recommendation are largely influenced by the physical resources that are available to you and the researchers that exist in your social network. Research is seldom a task done in isolation and modern AI research frequently requires access to substantial computational resources. For these reasons, measures of research output and letters of recommendations systemically disadvantage Black applicants and also feed back into the applicant pool in a way that amplifies existing bias. These components help to produce a system where it is hard to make meaningful changes along the lines of racial diversity and inclusion.  
    
    \par
    What can we do? The shortest answer is ``take risks.'' Right now, we screen for students that are similar to previous successful students in some manner. Admitting students whose experiences and achievements look differently than previous cohorts may seem risky; however, years of studying explore-exploit trade-offs should help us to more confidently take risks. If we continue to wait for a flood of Black candidates whose paper applications look identical to students that have been admitted for the last $K$ years, we are misunderstanding the fundamentals of systemic racism and how it contributes to what opportunities and achievements candidates with similar potentials and different backgrounds have access to.

    
    \item Who is mentored?
    \par
    If the goal of admissions is to select based upon research potential, the goal of mentorship is to develop research potential. We use mentorship as a catch-all term for early-career researchers (graduate and undergraduate) whom we help to train. A useful exercise may be to examine who you have written letters of recommendation for within the last $~5$ years. Does the diversity of this list reflect the diversity you would hope to see in the field? Unless we are very intentional, it is likely that our mentorship list is even less diverse than the department in which we work. The reasons for this are varied and may range from potential mentees' awareness of research opportunities, comfort reaching out to faculty, availability of research funding/ability to self-fund, and social connections with students within our labs. Ways to change this number may include actively reaching out to promising underrepresented students in our classrooms, participating in targeted summer REU programs, and publicizing transparent protocols on how interested students may get involved with our labs. If our campuses are not diverse enough to generate a robust pool of diverse students, we may establish closer relationships with other campuses. There are several reasons why the number of students we mentor may not reflect the diversity we seek to achieve, many of which have nothing to do with the qualification or talents of diverse students. Sometimes the work of allies is to put in extra effort to make themselves as accessible as possible.
    
    
    
    \item Who are your collaborators?
    \par
    
    Current students should be your primary source of collaboration, yet who you work with at other institutions matters greatly. While collaborating with familiar faces and institutions has its benefits, it also means that you are less likely to encounter diverse people through the course of your research. This affects not only whose work you are familiar with, but also who is familiar with your work. 
    

    \item What topics are emphasized and de-emphasized?
    \par
    What topics are valued in our community? Our community has done a tremendous job of simultaneously being driven by empirical performance of applications and avoiding consequential discussions on the implications of the applications that we focus on. 
    \par
    We write this piece in the midst of an intense international spotlight on anti-Black violence enacted and enabled through our policing systems. Artificial Intelligence systems have enabled more efficient and pervasive tracking, monitoring, criminalization, and repression of Black people. Historically, law enforcement organizations have deployed intense and technologically sophisticated surveillance campaigns with the goals of dismantling movements for the civil rights of Black people. We mention all of this to illustrate that while the AI community has been reticent to proffer regulations or even guidance on how AI technology should be ethically used, it has already been adapted and applied in various settings to exacerbate existing inequalities within our society. Diversifying our field will not remedy the harms that our systems have been agents in causing, but not addressing these harms may further drive marginalized groups from our community. This can reasonably be seen as being complicit in the harms our systems produce. 
    \par
    While work on fairness and ethical Artificial Intelligence exists, it has not received the same attention as more popular and controversial works such as facial recognition. The most-cited example o3f ethical AI that I have found has $ \sim1300$ citations \cite{barocas2016big}, while several facial recognition works, e.g., FaceNet, from the same time period have $\geq 3000$ \cite{parkhi2015deep, taigman2014deepface, schroff2015facenet}. Emphasising and centering research which deals with the societal implications of Artificial Intelligence is necessary to ensure that Artificial Intelligence has a positive impact on our societies.

    
    
    \item What schools/labs are feeders?
    \par
    Analyses of professors of Computer Science programs\cite{Huang:2014} indicate that 10 Computer Science programs are responsible for producing more than 50 percent of all CS faculty across the United States. 
    Disproportionate representation in these programs produces cascading effects that influence all other departments in the U.S. Sourcing graduate students or lab visitors primarily from institutions that have issues with retaining or attracting Black students to their programs further compounds the problem of under-representation because it allows these issues to spread to other schools. It is tempting to assume our problem with representation stems from a lack of suitable candidates. However, Black students making up $\geq 4\%$ of CS undergraduate degrees yet $\leq 1\%$ of PhDs, indicates that PhD programs suffer from a distinct drop-off that undergraduate diversity issues fail to explain. The reasons for this drop off are complex, but we believe one of them might be based on which schools serve as feeder schools into these PhD programs. Despite having $1/10^\text{th}$ the students, the 101 HBCUs produce more Black Computer Science Bachelor's degrees than the 115 R1 institutions in the U.S. PhD applications are designed to favor research experiences thus, the fact that such a large portion of Black CS undergraduates stem from institutions where research is not funded and/or prioritized as much as it is at R1 institutions puts Black applicants at a fundamental disadvantage. To conduct a quick self audit, you may look at the 11 HBCUs listed below, which graduate the largest numbers of Black CS students, and see how connected they are to your department \cite{Women:2018, Hbcu:2016}.
    \par
    \begin{itemize}
        \item North Carolina A \& T
        \item Southern University
        \item Norfolk State University
        \item Johnson C. Smith University
        \item Florida A \& M
        \item Alabama A \& M
        \item South Carolina State University
        \item Lane College
        \item Rust College
        \item University of Arkansas Pine Bluff
        \item Virginia State University
        \item Morehouse College
    \end{itemize}

    \item What companies are students recruited from/funneled into?
    \par

    Despite the aspirational nature of Artificial Intelligence's ideals, applications of AI have inflicted real harm on Black people. How you engage with companies that profit from anti-Black policies matters. Who do you take funding from? Who do you allow to recruit on campus? Who do you have informal relationships with? Where do your students end up working? I offer these as provocative questions and do not purport to offer prescriptive answers. However, I think it is important for our organizations to have these conversations and work on deciding our own ethical lines, if for no other reason than to mitigate the risk of scandal for money that we have accepted from nefarious sources.

    \item Who is hired?
    \par
    As we climb the ladder of academia, institutions become more and more risk-averse. Whereas there might be a willingness to take a chance on an undergraduate or summer visitor, this willingness drops when it comes to graduate students, post-docs, and especially faculty. If we want to truly move the needle on anti-Blackness in our community, the hiring and retention of Black professors must be addressed. The first stage of a research project is establishing relationships and, historically speaking, Black faculty have done a much better job of establishing relationships with Black students than other faculty. Fixing this discrepancy is not just about who is visible at the top, but about who is better positioned to succeed at all levels of your pipeline.
    \par
     Producing an equitable hiring process will require evaluating all aspects of the procedure with an eye for where biases may manifest. Are diversity efforts \textit{appropriately} valued with sufficient rigor and weighting? Are we drawing from representative pipelines and accounting for discrepancies in physical and social resources?

    
    
    \item How much do you pay?
    \par
    Systems that require financial ``sacrifice" in order to attain professional advancement are extremely effective at removing high achieving Black people from the leadership pipeline. The length of sacrifice and the drastic difference in earning potential and realized compensation of Computer Science PhD students makes this community one of the worst known examples of this discriminatory mechanism. 
    \par
    While discrepancies in graduate student pay affect all students, the 10x difference in median wealth between Black and White families means that Black students or potential students will be disproportionately impacted by a drop in income. This design choice, which has gone unaddressed for generations, creates a scenario in which a large number of qualified and interested Black students simply cannot afford to attend graduate school. Some schools and programs are aware that this opportunity cost is driving talented researchers away from academia and have developed quasi-official corporate partnerships that provide AI graduate students with supplemental incomes. However, these relationships are often opaque and invisible to students and communities without close insider connections. The unfortunate impact of this is that even if your program happens to be one that has made significant improvements to the compensation and quality of life of expected graduate students, potential Black students who would benefit the most from these changes are likely unaware of any improvements that you have made in this arena.
    

    
    
    \item What is the campus environment?
    \par
    
    Are the Black students on your campus adequately supported? This is a complicated question to answer and perhaps the best method of answering this question is by talking to your current and former Black students. A lot of students make it through graduate programs with traumatic experiences and unless there are Black professors, may not trust any faculty members enough to be candid about the problems in their environment. If and when you hear complaints from Black students, recognize that sharing these issues often constitutes a risk for these students and it is imperative to listen without becoming defensive. 
    

    \item What programs and resources are in-place?
    \par
    Systems that perpetuate anti-Black bias exist within the AI community just as strongly as they do in most other spaces. What resources have you committed to combating them? How has your lab, your department, your university, and larger community organizations in which you are a member addressed issues related to systemic anti-Blackness? Instead of proclaiming your beliefs and convictions, demonstrate them through investments and actions. Making significant changes in racial equity in our institutions is achievable but will require tangible effort from people other than those who are discriminated against.

\end{itemize}
\subsection{Graduate Students}
\begin{itemize}
    \item Who benefits from diversity?
    \par
    While substantial research has shown that diverse teams achieve better performance\cite{hunt2015diversity}, we reject this predatory view of diversity in which the worth of underrepresented people is tied to their value add to in-group members. We argue for combating anti-Blackness through the lens of justice. As such, all members of the community should be invested in developing a more inclusive and less discriminatory environment. 
    
    
    \item Who works towards diversity?
    \par
    Black graduate students (and faculty) shoulder an extraordinary amount of the burden of "diversifying" university campuses. One of the most impactful actions an ally can take is help to do this work. This participation will not only help to create a more diverse campus, but will also allow your Black peers to spend more time on their research and help them to attain successful graduate school careers. Understanding how to best assist in anti-Racism efforts may seem challenging or intimidating at first, but many of the necessary tasks are relatively simple and do not require an advanced degree in Black studies to result in meaningful contributions. Show up early, ask what you can do to support, volunteer to take over a non-leadership role. For example:  "Hey, I would like to help secure food for all the NSBE meetings this semester." While the current levels of attention on anti-racism might be fleeting, the magnitude of effort required for change is not; show up consistently and be willing to be lead.
    
    
    \item Who collaborates or studies with whom?
    \par
    Lack of social integration is a significant factor in disparate outcomes and experiences Black graduate students face in Academia. Lunches, dinners, happy hours, and game nights are activities that bear no direct relationship with technical merits but significantly contribute to the sense of safety and belonging that Black students in your program seek. As with most professions, social interactions often lead to professional opportunities. Within graduate school, study groups, research collaborations, and workshop organizers are often selected through a mix of social and academic ties. Reforming our social networks, as well as those of our entire community, is a formidable task, but while we are in the process of doing that we can actively acknowledge when biased social ties are influencing professional advancement and do our best to mitigate this. To state it frankly, put in the extra effort to know your Black colleagues' skills and interests and when opportunities present themselves that might align well with these students be sure that they are given fair consideration and not excluded because they are not in the right social groups.
    
    \item Which undergrads do you mentor?
    \par
    Who we invest our time in matters. The undergraduates that we mentor today will likely constitute the next generation of graduate students so it is imperative that we intentionally invest our time and mentorship capacity in diverse student populations. Oftentimes, the mechanisms of how to become involved in undergraduate research are hidden and subjective. One of the biggest levers we have to increase diversity in graduate school is to increase diversity in undergraduate research. To do this well, we should focus on both encouraging more underrepresented minorities to participate in undergraduate research and recruiting more graduate students to serve as mentors. 
    \par
    Additionally, it is necessary that we focus on finding funding for undergraduate researchers. Unpaid internships have a long history in the U.S. and one of their functions is to reinforce racial inequalities. We must work diligently at prohibiting this practice at our institutions. The ability to perform unpaid labor is often a privilege only afforded to those of substantial financial means. The following framing may help others more fully understand the role that compensation for undergraduate researchers plays in amplifying socio-economic inequality in our field. Let $C = U + P$ represent the total capacity for undergraduate mentorship within a department. While $U$ and $P$ represent the unpaid and paid researchers. The unpaid research slots, $U$, are reserved exclusively for high socio-economic students and the paid slots $P$ are to be split between students of all socio-economic backgrounds. Ignoring all other sources of systemic bias, this arrangement alone produce a system that significantly disadvantages Black students while serving as the primary pool for future graduate students. Calculating the true numbers for $P$ and $U$ in your institution should be a relatively straightforward process that can effectively illustrate just how skewed some of our current systems are.

    \item Who does the "devil's advocate" serve?
    \par
    Racism exists. It exists in our neighborhoods, in our departments, and in our labs. Conversations that question its existence are either intellectually lazy or conducted with ill intent. To have productive conversations and institute changes, we must avoid continually rehashing whether this well-documented phenomenon is a well-documented phenomenon. Just as a budget meeting filled with constant debate about the existence of currency is unproductive, so too is a discussion around anti-Blackness in which we refuse to name it or engage in meaningless debate over its existence. To address anti-Blackness, we cannot continue to indulge the whims of misinformed individuals or bad actors and offer them as much space as those working through active solutions.
    
\end{itemize}

\section{Discussion}
Through this work, we discuss several areas within the Artificial Intelligence community where systems perpetuate Anti-Blackness. This work falls woefully short of being comprehensive and intentionally does not discuss systems of racism within industry, conferences, or pipelines of capital. This work also does not address issues of sexism within our community, whose effects are compounded with racism to disproportionately impact Black women. The ultimate goal of this work is to assist in creating a more equitable Artificial Intelligence community whose broader impact includes reducing Anti-Blackness in our society. While not simple, we wholeheartedly believe that this is a tractable task and call upon our community to help make this happen.

\bibliographystyle{plain}
\bibliography{references}
\end{document}